\documentclass[sn-aps,iicol,pdflatex]{sn-jnl}

\usepackage{graphicx}%
\usepackage{multirow}%
\usepackage{amsmath,amssymb,amsfonts}%
\usepackage{amsthm}%
\usepackage{mathrsfs}%
\usepackage[title]{appendix}%
\usepackage{xcolor}%
\usepackage{textcomp}%
\usepackage{manyfoot}%
\usepackage{booktabs}%
\usepackage{algorithm}%
\usepackage{algorithmicx}%
\usepackage{algpseudocode}%
\usepackage{listings}%
\usepackage{graphicx}%
\usepackage{multirow}%
\usepackage{amsmath,amssymb,amsfonts}%
\usepackage{amsthm}%
\usepackage{mathrsfs}%
\usepackage[title]{appendix}%
\usepackage{xcolor}%
\usepackage{textcomp}%
\usepackage{manyfoot}%
\usepackage{booktabs}%
\usepackage{algorithm}%
\usepackage{algorithmicx}%
\usepackage{algpseudocode}%
\usepackage{listings}%
\usepackage{bm,physics}
\usepackage{amsmath,amssymb}
\usepackage{ulem, color}
\usepackage{amsmath}
\usepackage{natbib}
\bmdefine{\ba}{a}
\bmdefine{\bb}{b}
\bmdefine{\bx}{x}
\bmdefine{\by}{y}
\bmdefine{\bz}{z}
\bmdefine{\bn}{n}
\bmdefine{\bp}{p}
\newcommand{\BM}{\begin{pmatrix}}
\newcommand{\EM}{\end{pmatrix}}

\begin{document}

\title { Secondary bow  with    ripples 
in  $^{12}$C+$^{12}$C   rainbow scattering
}
\author*[1]{\fnm{S.} \sur{Ohkubo}}\email{ohkubo@rcnp.osaka-u.ac.jp}

\author[2]{\fnm{Y.} \sur{Hirabayashi}}

\affil*[1]{\orgdiv{Research Center for Nuclear Physics}, \orgname{Osaka University}, \orgaddress{\street{} \city{Ibaraki}, \postcode{567-0047}, \state{} \country{ Japan}}}

\affil[2]{\orgdiv{Information Initiative Center, Hokkaido University}, \orgname{Hokkaido University},\orgaddress{\street{} \city{Sapporo}, \postcode{060-0811 }, \state{} \country{Japan}}}

\pagenumbering{arabic}
\date{\today}
\pagenumbering{arabic}\pagenumbering{arabic}
\date{\today}
\pagenumbering{arabic}

\abstract{ We report,  {for the first time, the emergence of}  a secondary bow with ripples in $^{12}$C+$^{12}$C  nuclear rainbow scattering. This finding was achieved by studying the experimental angular distributions in $^{12}$C+$^{12}$C  scattering at incident energies  $E_L$= 240 and 300 MeV, utilizing an extended double-folding model. This model accurately describes all diagonal and off-diagonal coupling potentials derived from the microscopic wave functions for $^{12}$C. Although the observed angular distributions of rainbow scattering at large angles (approaching $90^\circ$) are complicated by the symmetrization of two identical bosonic nuclei, the Airy minimum, associated with a dynamically generated secondary bow with ripples, is clearly identified at approximately  77$^\circ$ for 240 MeV in the fall-off region of the primary nuclear rainbow. This {finding}, along with previous findings in the $^{16}$O+$^{12}$C and $^{13}$C+$^{12}$C systems, {reinforces} the concept of a secondary bow in nuclear rainbow scattering.
}


\maketitle

\section{Introduction}
\par
Rainbows have attracted humankind since long before the birth of science, and their mystery has been scientifically revealed since Descartes' time 
\cite{Descartes,Newton,Airy1838,Nussenzveig1977,Adam2002}. A microscopic nuclear rainbow discovered by Goldberg {\it et al} \cite{Goldberg1974} is a Newton's zero-order rainbow caused by refraction only \cite{Michel2002}, which was expected by Newton \cite{Newton} but not realized in the meteorological rainbow. Nuclear rainbows have been extensively studied 
\cite{Khoa2007,Michel2001,Ogloblin2003,Ohkubo2004A,Ohkubo2007,Hamada2013,Ohkubo2014,
Ohkubo2014B,Ohkubo2014C,Mackintosh2015,Ohkubo2015,Glukhov2007} and found to be very important in uniquely determining nucleus-nucleus interactions. 
The determined global deep potentials have been powerful in understanding nuclear cluster structures and nuclear rainbows in a unified way \cite{Michel1998,Ohkubo1999,Ohkubo2004B,Ohkubo2010,Hirabayashi2013}. Also, new concepts of prerainbow \cite{Michel2002}, dynamically refracted primary bow \cite{Ohkubo2016B}, dynamically generated secondary bow \cite{Ohkubo2014,Ohkubo2015B}, ripples superimposed on the Airy structure \cite{Ohkubo2014B}, and quasinuclear rainbows \cite{Ohkubo2024} have been developed.

\par
The existence of a secondary bow is not expected in principle in nuclear rainbow scattering caused by refraction due to an attractive nuclear potential. In fact, in the semiclassical theory of nuclear scattering \cite{Ford1959,Newton1966,Hodgson1978,Brink1985} within a mean-field nuclear potential, only one extremum (i.e., only one rainbow) is allowed in the deflection function. However, the existence of a secondary bow has been demonstrated in the asymmetric $^{16}$O+$^{12}$C \cite{Ohkubo2014} and $^{13}$C+$^{12}$C rainbow scattering \cite{Ohkubo2015B}. The secondary bow is generated dynamically \cite{Ohkubo2014,Ohkubo2015B} by a quantum coupling to the excited states in the classically forbidden dark side of the primary Newton's zero-order rainbow, which is caused by the astigmatism of a Luneburg-lens-like mean-field nuclear potential \cite{Michel2002,Luneburg1965,Ohkubo2016}. However, to the authors' best knowledge, no evidence of a secondary bow has been noticed for the thoroughly studied symmetric $^{12}$C+$^{12}$C system \cite{Buenerd1982,Bohlen1982,Bohlen1985,Brandan1982,Brandan1988,Brandan1990,%
Khoa1994,Brandan1997,Hassanain2008,Demyanova2010A,Demyanova2010B,%
Furumoto2012,Khoa2016,Hemmdan2021,Phuc2021}.

\par
The nuclear rainbow in $^{12}$C+$^{12}$C elastic scattering was first reported in Refs.~\cite{Brandan1982,Bohlen1982,Buenerd1982}. The experimental angular distribution measured at the incident energy $E_L$ =300 MeV by Bohlen {\it et al.} \cite{Bohlen1982} was reproduced well up to $\theta_{_{{\rm c.m.}}}$=63$^\circ$, except for the rise at the largest angle of 70.4$^\circ$, in the folding model with a coupling to the  $2^+$ (4.44 MeV) of $^{12}$C. The nuclear rainbow angle  $\theta_{\rm c.m.}^R$$\approx$56$^\circ$
  was determined from the deflection function. Bohlen {\it et al.} \cite{Bohlen1985} also measured the angular distribution at  $E_L$=240 MeV up to  $\theta_{{\rm c.m.}}$=60$^\circ$. They observed a clear first-order Airy minimum (A1) at $\theta_{{\rm c.m.}}$$\approx$40$^\circ$  followed by the fall-off of the cross sections in the dark side of the nuclear rainbow. Demyanova {\it et al.} \cite{Demyanova2010A,Demyanova2010B} extended the measurement of the angular distribution at $E_L$=240 MeV  to a much larger angle of 93$^\circ$, which showed that the angular distribution does not fall monotonically beyond 60$^\circ$, similar to the case at 300 MeV.

 \par
  Thus, the experimental angular distributions in rainbow scattering for $^{12}$C+$^{12}$C at $E_L$ =240-300 MeV \cite{Bohlen1982,Bohlen1985,Demyanova2010A,Demyanova2010B} have not been reproduced at large angles beyond 60$^\circ$ toward 90$^\circ$ in intensive calculations performed on purpose \cite{Bohlen1985,Demyanova2010A,Demyanova2010B,Khoa2016}. Despite extensive studies of $^{12}$C+$^{12}$C scattering \cite{Khoa1994,Hassanain2008,Demyanova2010A,Demyanova2010B,Furumoto2012,Khoa2016,Hemmdan2021,Phuc2021}, the discrepancy between theory and experiment remains unsolved. Obviously, $^{12}$C+$^{12}$C rainbow scattering cannot be understood within the frame of the ordinary theory of nuclear rainbow \cite{Ford1959,Newton1966,Hodgson1978,Brink1985}. No attention has been paid to the physical meaning and the origin of the anomalous structure of the angular distributions beyond 60$^\circ$ in the fall-off region.
  
  \par
  The purpose of this paper is to report, for the first time, {the emergence} of a secondary bow  with ripples in $^{12}$C+$^{12}$C rainbow scattering at both $E_L$ =240 and 300 MeV. We investigate the angular distribution of $^{12}$C+$^{12}$C rainbow scattering using the coupled channels (CC) method with an extended double folding potential derived from realistic microscopic wave functions for $^{12}$C and a density-dependent effective two-body force. 
   This third finding, alongside the $^{16}$O+$^{12}$C and $^{13}$C+$^{12}$C systems, {reinforces} the concept of secondary bow in nuclear rainbow scattering.
   The mechanism of the secondary bow is investigated.

  \section{The extended double folding model}
\par
We study rainbow scattering for the $^{12}$C+$^{12}$C system with an extended double folding (EDF) model that describes all the diagonal and off-diagonal coupling potentials derived from the microscopic realistic wave functions for $^{12}$C using a density-dependent effective nucleon-nucleon force. The diagonal and coupling potentials for the $^{12}$C+$^{12}$C system are calculated using the EDF model as follows:
\begin{eqnarray}
\lefteqn{V_{ij,kl}({\bf R}) =
\int \rho_{ij}^{\rm (^{12}C)} ({\bf r}_{1})\;
     \rho_{kl}^{\rm (^{12}C)} ({\bf r}_{2})} \nonumber\\
&& \times v_{\it NN} (E,\rho,{\bf r}_{1} + {\bf R} - {\bf r}_{2})\;
{\it d}{\bf r}_{1} {\it d}{\bf r}_{2} ,
\end{eqnarray}
\noindent where
$\rho_{ij}^{\rm (^{12}C)} ({\bf r})$ represents the diagonal ($i=j$) or transition ($i\neq j$)
 nucleon density of $^{12}$C which is calculated using the microscopic three $\alpha$ cluster model in the resonating group method \cite{Kamimura1981}. 
This model well reproduces the $\alpha$ cluster and shell-like structures of $^{12}$C, and the wave functions have been checked against much experimental data, including charge form factors and  electric transition probabilities \cite{Kamimura1981}.
In the calculations, we take into account the excitation of the  $2^+$, $3^-$ (9.64 MeV) and $4^+$ (14.08 MeV) states of $^{12}$C. For the effective nucleon-nucleon interaction ($v_{\rm NN}$), we use the DDM3Y-FR interaction \cite{Kobos1982,Kobos1984,Bertsch1977}, which accounts for the finite-range nucleon exchange effect \cite{Khoa1994}. We introduce the normalization factor $N_R$ 
  \cite{Brandan1997} for the real double folding potential. An imaginary potential with a Woods-Saxon volume-type (nondeformed) form factor is phenomenologically introduced to account for the effect of absorption due to other channels.

 \section{Emergence of a secondary nuclear  rainbow }
\par
\begin{figure}  [t]
\includegraphics[keepaspectratio,width=7.0cm] {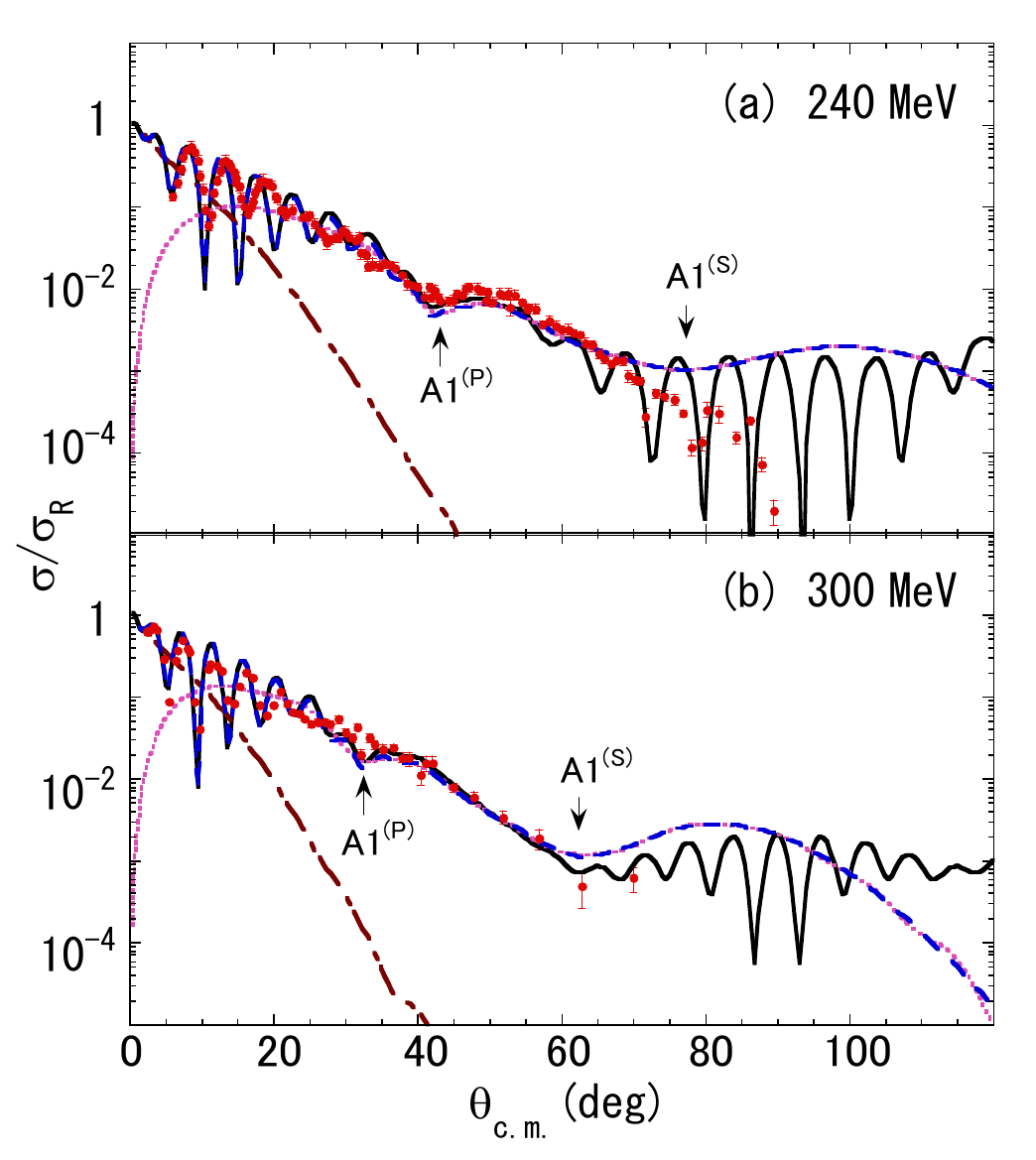}
 \protect\caption{\label{fig1} {
Angular distributions in $^{12}$C+$^{12}$C scattering at (a)  $E_L$ = 240 MeV and (b) 300 MeV, calculated  with the six-channel couplings, are displayed (as a ratio to Rutherford scattering)  as black solid lines in comparison with the experimental data (points) from Refs.~\cite{Bohlen1982,Bohlen1985,Demyanova2010A,Demyanova2010B}. The results without symmetrization of two bosonic  identical nuclei are displayed by blue dashed lines. The calculated farside  and nearside components are given by pink dotted  lines and brown dash-dotted lines, respectively. $A1^{(P)}$ and $A1^{(S)}$ stand for the Airy minimum of the primary nuclear rainbow and the secondary bow, respectively
}
}
\end{figure}

In Fig.~\ref{fig1}, the angular distributions for elastic $^{12}$C+$^{12}$C scattering at  $E_L$= 240 and 300 MeV, calculated using the CC method with the six-channel couplings  of ($^{12}$C(${ I}^\pi$), $^{12}$C(${ J}^\pi$)) =  ($0^+$,  $0^+$), ($0^+$,  $2^+$),  ($0^+$,  $3^-$), ($0^+$,  $4^+$), ($2^+$,  $2^+$), and ($2^+$ ,  $4^+$), 
  are displayed in comparison with the experimental data \cite{Bohlen1982,Bohlen1985,Demyanova2010A,Demyanova2010B}.
  We used $N_R$ = 1.13 and 1.20 for the real potential at 240 and 300 MeV, respectively.
  For the imaginary potential the strength parameters $W$=18 and 19 MeV for  $E_L$=240 and 300 MeV, respectively, were found to fit the data with fixed radius and diffuseness parameters $R_W$=5.6 fm and  $a_W$=0.7 fm.

 The calculated symmetrized angular distributions (solid lines) show good agreement with the experimental data up to 70$^\circ$.  Our calculations accurately reproduce the Airy minima at both 240 MeV and 300 MeV. 
 At 240 MeV, our calculations also reproduce the oscillatory ripple pattern for angles beyond 70$^\circ$, although the magnitude is somewhat larger than the experimental data. 
 We also successfully reproduce the angular distribution at 300 MeV, including the final data point at 70$^\circ$, where the cross section appears to begin to rise.
It is highly desirable to measure the angular distribution at large angles at 300 MeV to confirm the predicted ripple oscillations.
These good fits to the data at 240 MeV and 300 MeV are much better than the latest calculations in Refs. ~\cite{Khoa2016,Phuc2021}.

\par
We note the rise of the cross sections both at 240 and 300 MeV beyond $\theta_{\rm c.m.}$$\approx$$60^\circ$ in the fall-off region of the primary nuclear rainbow and the appearance of a broad bump toward  $90^\circ$  in the calculated unsymmetrized angular distributions shown as dashed blue lines. 
This resembles the secondary bow observed in the asymmetric $^{13}$C+$^{12}$C scattering, where a bump of the secondary bow appears at $\theta_{{\rm c.m.}}$$\approx$$70^\circ$ (Fig. 1 of Ref. \cite{Ohkubo2015B}).

To analyze this, we decomposed the calculated angular distributions into farside and nearside components, following the powerful prescription of Refs. \cite{Fuller1975,McVoy1984}. This method has been very effective  in studying the Airy minimum in rainbow scattering, including two  bosonic identical nuclei $^{12}$C+$^{12}$C \cite{McVoy1992,Michel2004} and $^{16}$O+$^{16}$O \cite{Nicoli1999,Michel2001}. The farside component is shown by the pink dotted lines. We find that the enhanced cross sections of the bump in the angular distributions beyond $\theta_{\rm c.m.}$$\approx$$60^\circ$ are due to refractive farside scattering.
The first order Airy minimum $A1^{(P)}$ of the primary nuclear rainbow is determined at  43$^\circ$ for $E_L$=240 MeV and at 33$^\circ$ for 300 MeV, respectively. The Airy minimum of the secondary bow $A1^{(S)}$ is located at $\theta_{\rm c.m.}$=77$^\circ$ for 240 MeV and 63$^\circ$ for 300 MeV, respectively.

\par
We note that the experimental angular distribution beyond $\sim$70$^\circ$ at 240 MeV shows oscillations, and the rise of the last data point of the angular distribution at 300 MeV is an indication of the start of the secondary bow.
 The present calculations (solid lines) reproduce these characteristic oscillatory features of the angular distributions at large angles toward 90$^\circ$. The oscillations are due to symmetrization of identical two bosonic nuclei, which causes interference between the scattering amplitudes at $\theta_{\rm c.m.}$ and 180-$\theta_{\rm c.m.}$.
The interference breaks up the bright bump of the secondary bow, producing the ripples, which makes it difficult to intuitively notice the existence of the bright secondary bow in symmetric $^{12}$C+$^{12}$C rainbow scattering, unlike in the asymmetric $^{16}$O+$^{12}$C and $^{13}$C+$^{12}$C systems.

 \begin{figure}[t!]
\includegraphics[keepaspectratio,width=7.0cm] {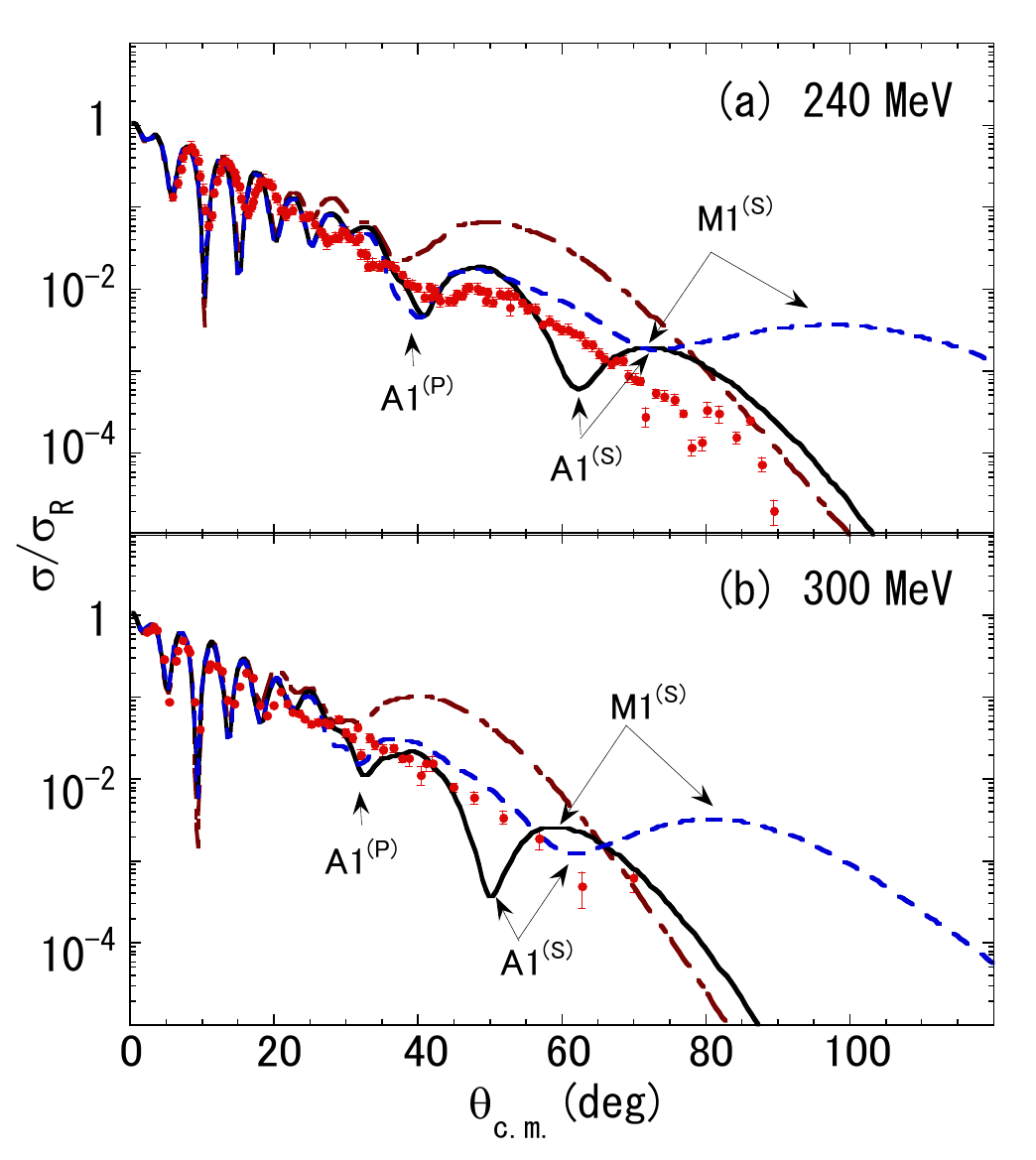}
\protect\caption{\label{fig.2} {
 The angular distributions for $^{12}$C+$^{12}$C scattering calculated without symmetrization at (a)  $E_L$=240 and (b) 300 MeV are displayed (as a ratio to Rutherford scattering) for: single-channel 
calculations (dash-dotted lines), two-channel calculations 
(solid lines),  and three-channel calculations
  (dashed lines). These are compared with the experimental data (points) from Refs.\cite{Bohlen1985,Demyanova2010A,Demyanova2010B}.
    $A1^{(P)}$ stands for the Airy minimum of the primary nuclear rainbow, while
     $A1^{(S)}$  and $M1^{(S)}$  represent the Airy minimum and maximum of the secondary bow, respectively
   }
}
\end{figure} 

\section{Origin of a secondary bow and ripples}
\par
To investigate which coupling is responsible for the generation of the secondary bow with the Airy minimum  $A1^{(S)}$, we display in Fig.~\ref{fig.2} the angular distributions calculated under three different coupling scenarios: (1) in the single channel, (2) with two-channel coupling including ($0^+$, $0^+$) and ($0^+$, $2^+$), and (3) with three-channel coupling including  ($0^+$, $0^+$), ($0^+$, $2^+$),  and  ($2^+$, $2^+$) channels.
It is clear that the single-channel calculations produce only the first-order Airy minimum $A1^{(P)}$ of the primary nuclear rainbow, and the angular distributions in the dark side fall monotonically at large angles toward 90$^\circ$. We observe that $A1^{(S)}$ and the bright bump of the Airy maximum $M1^{(S)}$ of the secondary bow, appearing before the fall-off, are already created by the two-channel coupling.
By including mutual excitation to the  $2^+$  state in the three-channel coupling, $A1^{(S)}$ and     $M1^{(S)}$  shift to larger angles, becoming closer to the calculations with the six-channel coupling shown in Fig.~\ref{fig1}. Thus, coupling to the $2^+$ state with reorientation \cite{Ohkubo2015B} is essential for creating the secondary bow.

\begin{figure}[t]
\includegraphics[keepaspectratio,width=7.0cm] {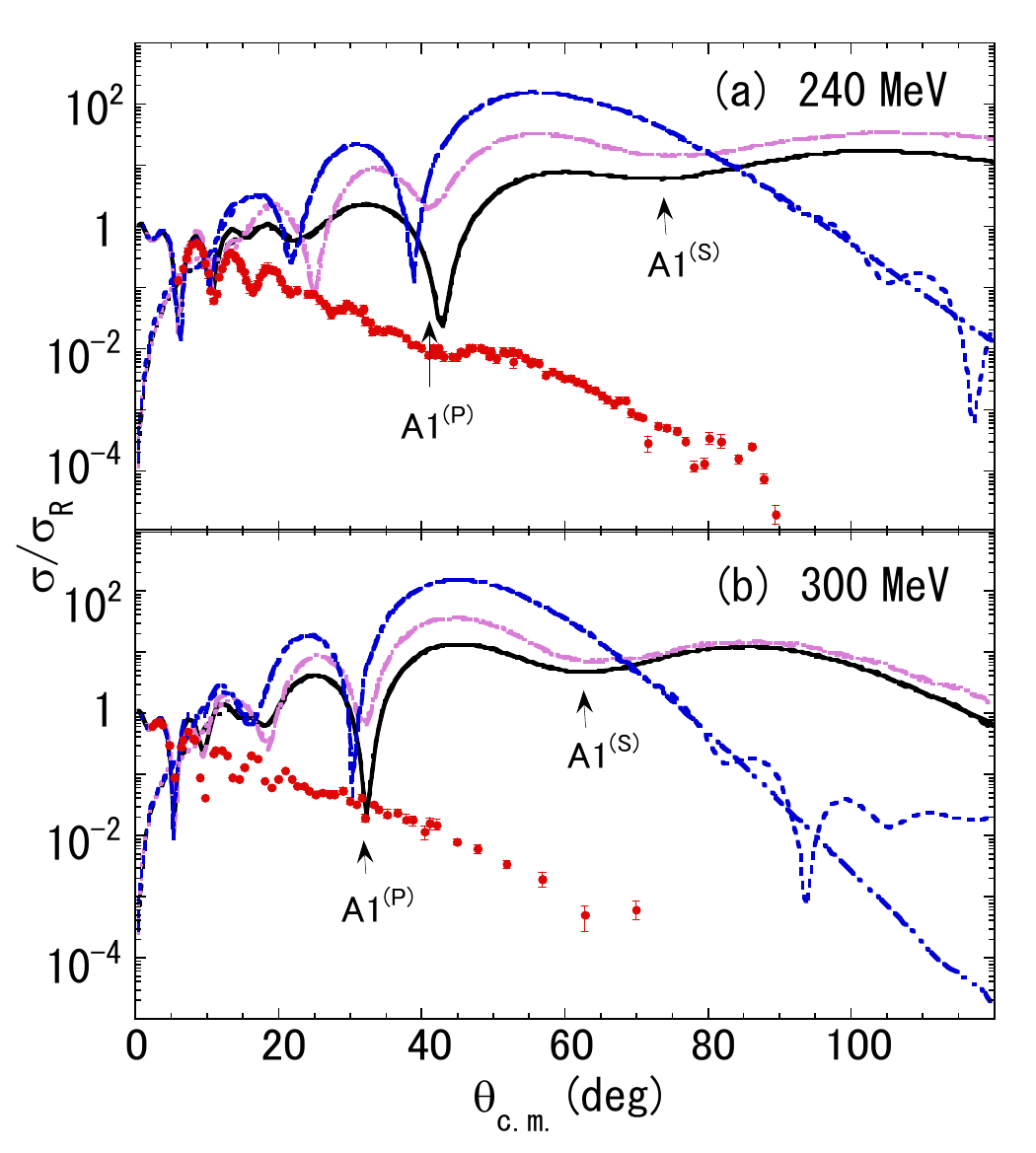}
 \protect\caption{\label{figNFW=0} {
  The unsymmetrized total (farside) angular distributions for  $^{12}$C+$^{12}$C scattering at  (a) 240 MeV and (b) 300 MeV,  calculated with   $W$=0,  are displayed (as a ratio to Rutherford scattering) by the blue dash-dot-dot lines (blue dashed lines) for the single-channel coupling, pink dashed lines (pink dotted lines) for the three-channel coupling, and black solid lines (black dot-dash lines) for the six-channel  coupling.
 The farside components are indistinguishable from the total angular distributions in the rainbow region including  $A1^{(P)}$ and  $A1^{(S)}$.
 The experimental data (points) are from Refs.~\cite{Bohlen1982, Bohlen1985, Demyanova2010A, Demyanova2010B}
}
}
\end{figure}

\par
Fig. \ref{figNFW=0} clearly shows the Airy minima  $A1^{(P)}$ and   $A1^{(S)}$  in calculations performed with a reduced imaginary potential $W$=0. In the single-channel calculations (blue dash-dot-dot lines), only $A1^{(P)}$ is distinctly observed at 240 and 300 MeV. For the three-channel calculations, the primary rainbow's fall-off is replaced by enhanced cross sections, forming a bright region of the secondary bow with $A1^{(S)}$,  while  the position of $A1^{(P)}$ is only slightly shifted backward compared to calculations without couplings. The six-channel calculations (solid lines) exhibit $A1^{(P)}$ at  43$^\circ$ and $A1^{(S)}$ at 77$^\circ$ for 240 MeV, and $A1^{(P)}$ at 33$^\circ$ and  $A1^{(S)}$ at  63$^\circ$ for 300 MeV. As seen in Fig. \ref{figNFW=0}, the farside components are indistinguishable from the total angular distributions in the regions of $A1^{(P)}$ and $A1^{(S)}$ of the Airy structure. This observation indicates that the secondary bow with $A1^{(S)}$ originates from refractive farside scattering.

\begin{figure}[t!]
\includegraphics[keepaspectratio,width=7.0cm] {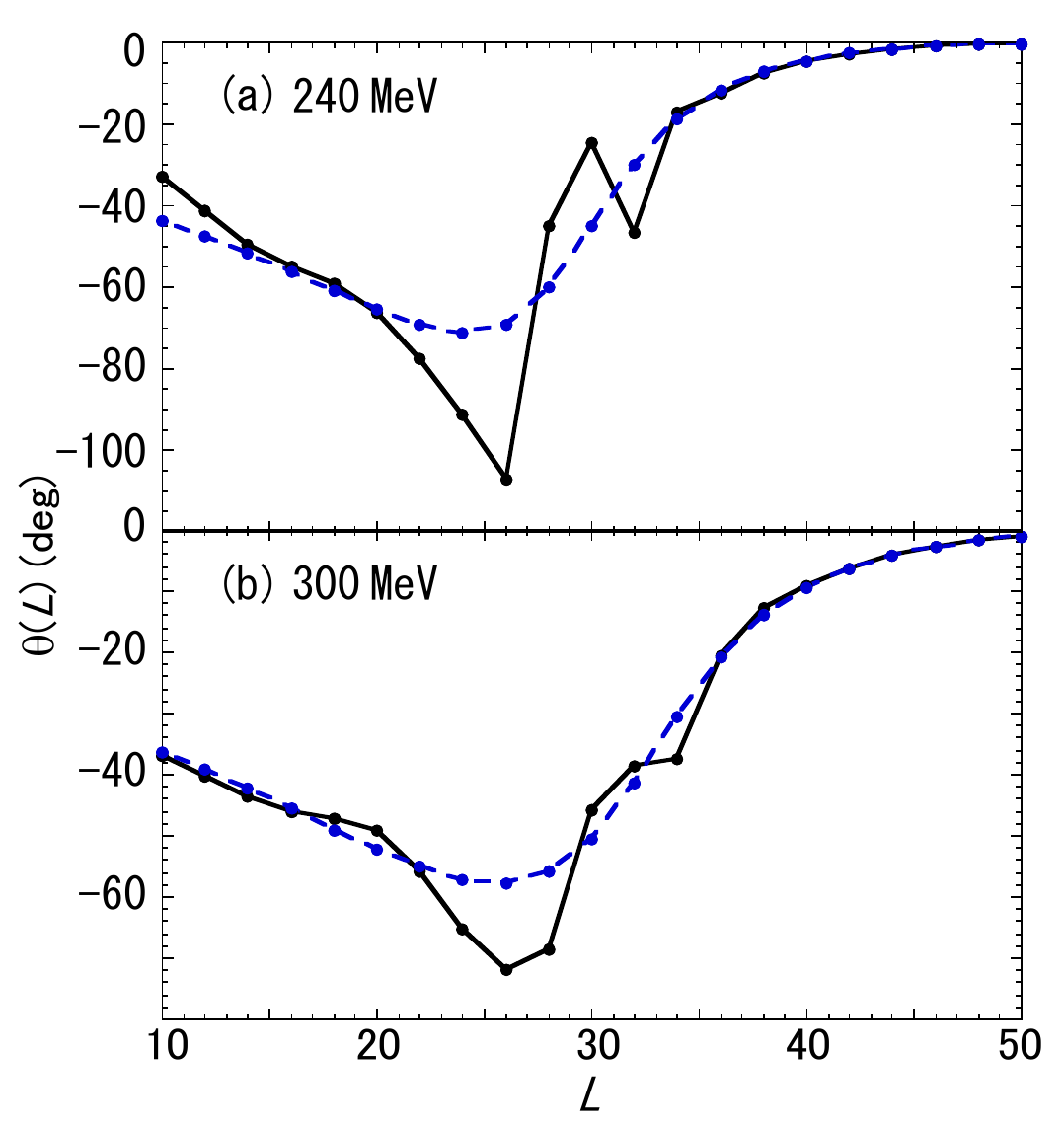}
\protect\caption{\label{deflection} {
The deflection functions for $^{12}$C+$^{12}$C scattering, calculated with $W$=0 in  the 
six-channel coupling   (solid lines) and in  the single-channel
  (dashed lines),  are displayed at 240 and 300 MeV.
  The line is to guide the eye
   }
}
\end{figure}
\par
In Fig.~\ref{deflection}, while the single-channel calculations show one extremum in the deflection functions, corresponding to the primary nuclear rainbow angle, the CC calculations show more than one extremum. The secondary bow, which corresponds to the second minimum, is dynamically caused by quantum coupling to the internal structure of the $^{12}$C nuclei. While Newton's zero-order primary nuclear rainbow is caused by the radially monotonic mean field potential and involves no internal structure, the secondary bow is created due to the dynamical polarization potential with undulations \cite{Mackintosh2015}. The undulations of the deflection function arise from the undulations of the polarization potential, which appears even in the intermediate radial region due to the coupling to the $2^+$ state of $^{12}$C. This explains why it was not possible to reproduce the angular distributions for $^{12}$C+$^{12}$C rainbow scattering up to large angles toward 90$^\circ$ with the mean field potential alone, without the dynamically generated polarization potential with undulations.
In Fig.~\ref{deflection}, we note that partial waves (trajectories) with larger $L$ values than those for the primary rainbow are involved. This means that the secondary bow emerges in the outer region of the nucleus where the inelastic scattering to the $2^+$  
  state occurs, and the refractive scattering of the primary rainbow is strongly damped.
 
\begin{figure}[t!]
\includegraphics[keepaspectratio,width=7.4cm] {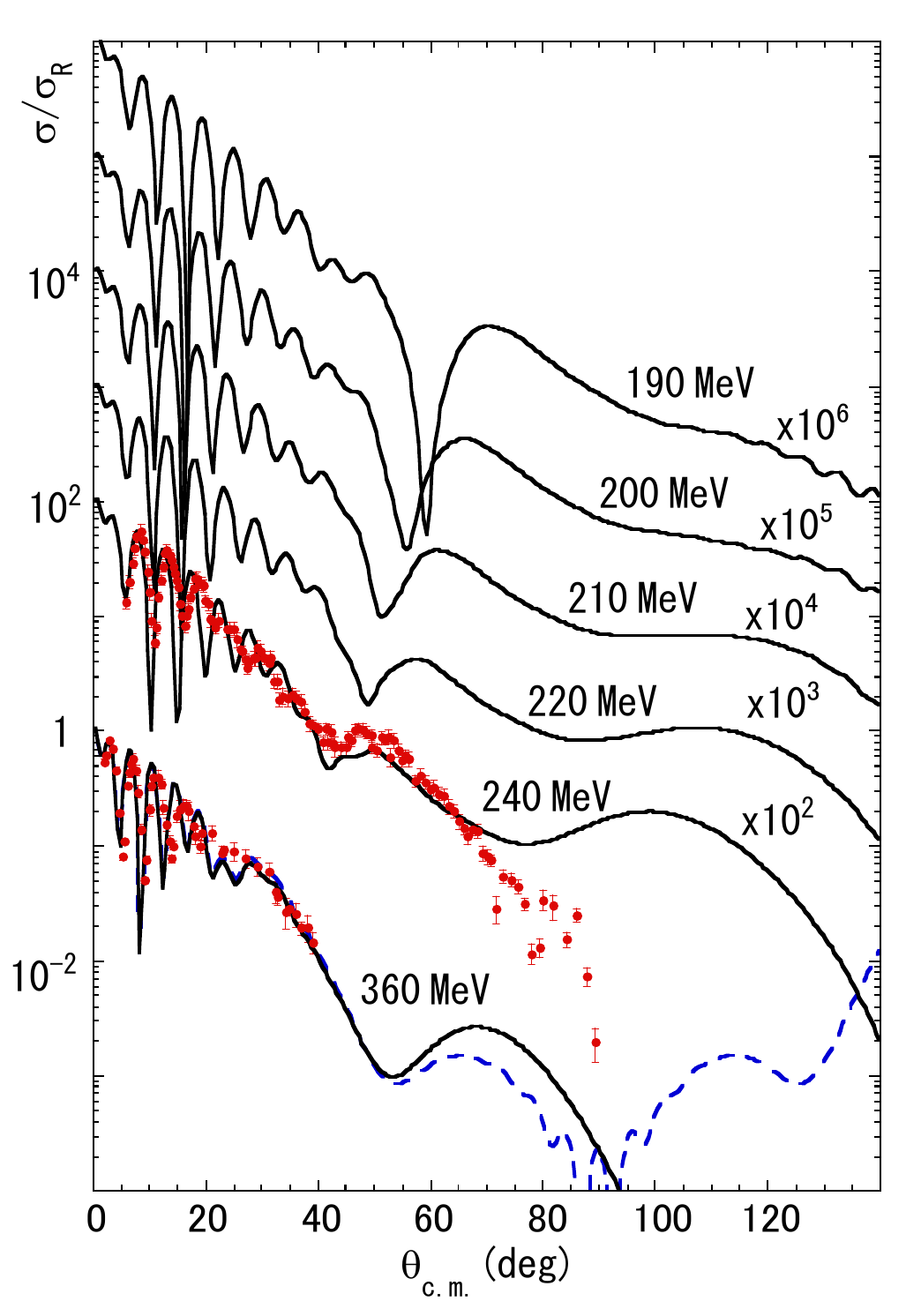}
\protect\caption{\label{fig5growth} {
Angular distributions calculated without symmetrization at energies above 190 MeV with the six-channel coupling are displayed (as a ratio to Rutherford scattering) as solid lines. At 360 MeV, dashed lines represent the symmetrized calculated angular distribution in comparison with the experimental data  (points) from Ref.~\cite{Buenerd1984}
}
}
\end{figure}

\par
Since a secondary bow is not seen in the experimental angular distributions at 139.5 and 158.8 MeV in Refs.~\cite{Kubono1983,Kubono1985}, we investigate at what energy the secondary bow emerges. In Fig.~\ref{fig5growth}, the angular distributions calculated with the six-channel coupling at $E_L$=190-360 MeV, using the potential at 240 MeV (except $N_R$=1.2 for 360 MeV), are displayed.
At 200 MeV, the fall-off in the dark side of the primary rainbow, with  $A1^{(P)}$ at 56$^\circ$ and  $M1^{(P)}$ at 66$^\circ$ in the bright side, continues toward large angles, showing no indication of a secondary bow. At 210 MeV, this fall-off stops, and a plateau appears at around 90$^\circ$, followed by a new fall-off beyond 120$^\circ$. At 220 MeV, we observe a clear minimum,  $A1^{(S)}$, at 88$^\circ$, and the bump of the secondary bow, $M1^{(S)}$, at 108$^\circ$. 
 Thus, we find that the secondary bow starts to emerge at around 210 MeV and develops as the incident energy increases.

The calculation accurately reproduces the experimental data at 360 MeV from Ref.~\cite{Buenerd1984}. The calculation places the  $A1^{(P)}$ at around 25$^\circ$ and predicts the secondary bow with $A1^{(S)}$ at around 54$^\circ$ and  M1$^{(S)}$  at around 70$^\circ$. The secondary bow is predicted to persist at least as high as 360 MeV. 

At this higher energy, ripples from symmetrization, which characterize the secondary rainbow of the symmetric $^{12}$C+$^{12}$C system and can obscure the original M1$^{(S)}$ bump, appear near 90$^\circ$. This makes it difficult to recognize the existence of a secondary rainbow in experimental data.
However, these ripples will disappear in the secondary rainbow for the $^{12}$C+$_\Lambda^{13}$C system, because the $\Lambda$ hyperon breaks bosonic symmetrization. 
In fact, the M1$^{(S)}$ bump of the secondary rainbow, now without ripples, has been confirmed in experimental data from $^{13}$C + $^{12}$C scattering, where an extra neutron similarly breaks bosonic symmetrization (see Fig. 1 of Ref.~\cite{Ohkubo2015B}).
It is  worth mentioning that ripples arise from the symmetrization of two bosonic identical nuclei \cite{Satchler1983} in refractive scattering, not solely from the symmetrization of either the farside or nearside scattering amplitude.

\begin{figure}[t!]
\includegraphics[keepaspectratio,width=8.0cm] {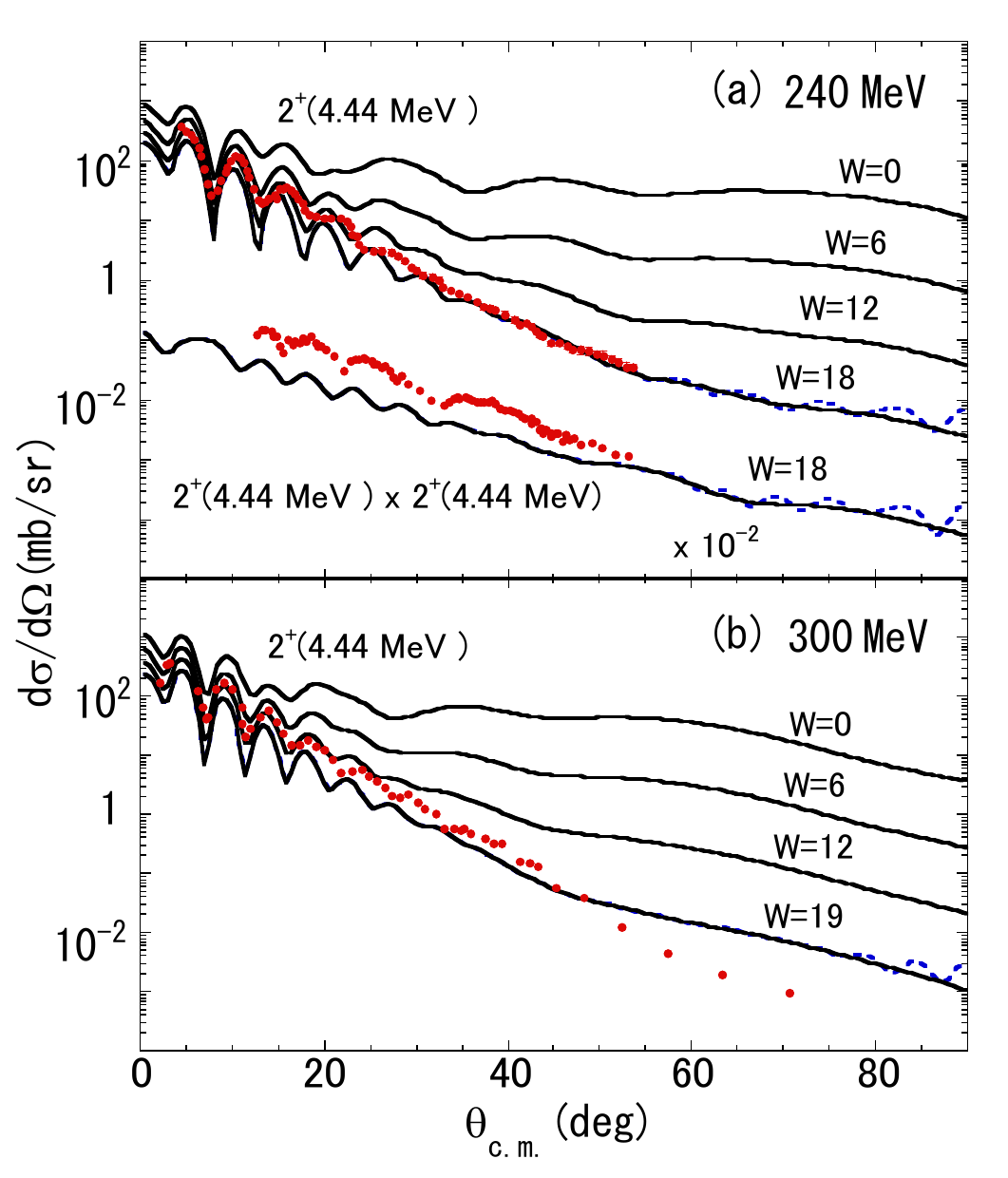}
 \protect\caption{\label{fig6inelastic} {
Angular distributions for inelastic $^{12}$C+$^{12}$C scattering with excitation to the $2^+$  state at (a) $E_L$=240 and (b) 300 MeV, calculated
 with the six-channel coupling (solid lines), are displayed in comparison with the experimental data (points) from Refs. \cite{Bohlen1982,Bohlen1985,Demyanova2010A,Demyanova2010B}
 }
}
\end{figure}
\par
The reason why the secondary bow is generated above $E_L$=210 MeV is that the contribution to the elastic scattering cross sections from channel couplings, predominantly from the $2^+$ state of $^{12}$C, increases relative to the cross sections solely due to the elastic channel. The latter decreases exponentially at large angles in the fall-off region on the dark side of the primary nuclear rainbow. The inelastic scattering to the $2^+$ state, which is shown in  Fig.~\ref{fig6inelastic}, is very strong, and its cross sections even exceed those of other inelastic and even elastic channels in the rainbow region.
As the incident energy increases, the  $A1^{(P)}$ of the primary rainbow shifts to forward angles, which causes the fall-off to shift forward. Thus, the cross sections in the fall-off region of the primary rainbow decrease rapidly as energy increases, which enhances the prominence of the bump of the secondary bow, $M1^{(S)}$. Thus,  the secondary bow is generated dynamically   at energies above 210 MeV.

Finally, we mention that  the  long-standing problem concerning the Airy minima and Airy elephants in $^{12}$C+$^{12}$C scattering has  been resolved after decades of concern by recognizing the existence of a dynamically generated secondary rainbow
 \cite{Ohkubo2025}.

 \section{Summary}
\par
We have, for the first time, shown {the emergence of}  a secondary bow with ripples in the angular distributions of symmetric $^{12}$C+$^{12}$C rainbow scattering at $E_L$=240 and 300 MeV. 
To achieve this, we utilized an extended double-folding model with coupled-channel calculations, incorporating diagonal and off-diagonal potentials derived from microscopic wave functions of $^{12}$C. This {finding}, alongside  the asymmetric systems $^{16}$O+$^{12}$C and $^{13}$C+$^{12}$C, {reinforces} the  concept of a secondary bow in nuclear rainbow scattering. 
We suggest that a secondary bow exists widely in nuclear rainbow scattering involving $^{12}$C, such as $^{12}$C+$^{13}$N, $^{12}$C+$^{14}$C, $^{12}$C+$^{15}$C, $^{12}$C+$_\Lambda^{13}$C, and $^{12}$C+$_\Lambda^{17}$O.

 \begin{itemize}
\item Funding
No funding.
\item Conflict of interest/Competing interests 
Not applicable
\item Consent for publication
The authors agree with publication of the manuscript in the traditional publishing model.
\item Data availability Statement
This manuscript has no associated data or the data will not be deposited. [Authors’ comment: This is a theoretical study using  published data, and all data information is properly referenced.]
\item Code availability 
Code/Software sharing not applicable to this article as
this manuscript has no associated code/software.

\end{itemize}

{}
\end{document}